\documentclass[fleqn,10pt]{wlscirep}
\usepackage{graphicx}
\usepackage{amsmath}
\usepackage{amssymb}
\usepackage{subcaption}

\newcommand{\beq}{\begin{equation}}
\newcommand{\eeq}{\end{equation}}

\title{A control analysis perspective on Katz centrality}

\author[1,*]{Kieran J. Sharkey}
\affil[1]{Department of Mathematical Sciences, University of Liverpool, Liverpool, L69 7ZL, UK}

\affil[*]{kjs@liverpool.ac.uk}

\begin{abstract}
Methods for efficiently controlling dynamics propagated on networks are usually based on identifying the most influential nodes. Knowledge of these nodes can be used for the targeted control of dynamics such as epidemics, or for modifying biochemical pathways relating to diseases. Similarly they are valuable for identifying points of failure to increase network resilience in, for example, social support networks and logistics networks. Many measures, often termed `centrality', have been constructed to achieve these aims. Here we consider Katz centrality and provide a new interpretation as a steady-state solution to continuous-time dynamics. This enables us to implement a sensitivity analysis which is similar to metabolic control analysis used in the analysis of biochemical pathways. The results yield a centrality which quantifies, for each node, the net impact of its absence from the network. It also has the desirable property of requiring a node with a high centrality to play a central role in propagating the dynamics of the system by having the capacity to both receive flux from others and then to pass it on. This new perspective on Katz centrality is important for a more comprehensive analysis of directed networks.
\end{abstract}
\begin{document}

\flushbottom
\maketitle

\thispagestyle{empty}

\section*{Introduction}
\label{S1}
Ranking nodes with respect to their degree often provides valuable information on their relative importance, but it can also neglect essential factors and, by definition, it cannot distinguish between nodes of the same degree. One of the major deficiencies of degree centrality is that while it counts the number of neighbours of a node, it does not account for how central or important those neighbours are. Centrality measures that directly address this deficiency include Katz centrality \cite{Katz}, eigenvector centrality \cite{Bonacich}, and the centrality known as PageRank which partly underpins the Google search engine \cite{BrinPage}. We briefly summarise these here.

We consider a network represented by an $n$ by $n$ non-negative matrix $A$ such that $A_{ij}\neq 0$ represents a connection (arc) from node $j$ to node $i$, and potentially $A$ could be weighted and directed. Katz centrality \cite{Katz} is the earliest of these measures and can be defined by
\beq
{\bf r}={\bf 1}+aA{\bf r},
\label{centrality_equation}
\eeq
where ${\bf r}=[r_1,r_2,...,r_n]^T$ denotes Katz centrality at each node, ${\bf 1}$ is an $n\times 1$ column vector of ones and $0< a<1/\rho(A)$ where $\rho(A)$ is the spectral radius of matrix $A$ (see methods). This has solution
\beq
{\bf r}=(I-aA)^{-1}{\bf 1}=M{\bf 1},
\label{cent_eq}
\eeq
where $M=\left (I-aA\right )^{-1}$ and $I$ is the $n$ by $n$ identity matrix. This definition is slightly different to the original paper by Katz\cite{Katz}, but it results in the same ranking of nodes. It can be interpreted as a sum of two contributions. One is an intrinsic centrality that each node has (first term on the right of (\ref{centrality_equation})) and the other is the centrality passed to it in proportion to how important its neighbours are (second term on the right of (\ref{centrality_equation})). PageRank \cite{BrinPage} can be viewed as a variant of Katz centrality\cite{Newman} in which the centrality contributed by a node to its neighbours is treated as a limited resource which is distributed evenly among them. Experience by Google suggests that this modification makes a better representation of the relative importance of internet pages.

Eigenvector centrality \cite{Bonacich} can be motivated in a similar way to Katz centrality and is defined as the principal right-eigenvector of $A$, which we shall denote by ${\bf u}$. Although there are clear similarities between (\ref{centrality_equation}) and the eigenvalue equation, and similarities with the motivations behind the two measures, they are mathematically distinct.

Complications arise in all of these centralities when we consider directed networks. In the context of eigenvector centrality, Kleinberg \cite{Kleinberg} proposed a resolution by using two quantities to characterise a directed network. These are termed `hubs' and `authorities' where each node has a measure of the extent to which it is behaving like a hub (receiving centrality from its neighbours) and the extent to which it is behaving like an authority (passing centrality to its neighbours). The hub quantifier, which we shall denote by ${\bf h}$, is defined by the principal right-eigenvector of matrix $A^TA$. The authority quantifier, which we shall denote by ${\bf a}$, is defined as the principal right-eigenvector of $AA^T$. For undirected networks, this distinction is unimportant and both reduce to standard eigenvector centrality. 

On directed networks, Katz centrality, PageRank and eigenvector centrality favour hub-like nodes. This is evident from the form of (\ref{centrality_equation}) and from the eigenvalue equation where centrality is passed to a node from its inward-pointing neighbours. However, we argue here that a genuinely central node, in the sense that it is central to the ongoing dynamics, should be able to receive flux from others and be able to pass this on. In this sense, pure source or pure sink nodes are peripheral to sustaining the ongoing dynamics and so are not central, even though a sink node could attain a high centrality from the form of (\ref{cent_eq}). 

We first recast Katz centrality as a solution to a continuous-time dynamical system. Then, by investigating the idea that the importance or centrality of an individual to this system is determined by the net impact of their removal, with everything else remaining the same, we quantify the net worth or effect of that individual, both in terms of their intrinsic value as well as by the impact that they have on others. By making a linear approximation to this process of node deletion we derive a centrality measure that automatically accounts for both the hub-like and the authority-like properties of nodes and gives a different perspective on Katz centrality.

\section*{Methods}
\subsection*{A linear continuous-time dynamical system}
We can interpret centrality of the form (\ref{cent_eq}) as the unique steady state of the following linear continuous-time dynamical system:
\beq
\frac{d{\bf x}}{dt}={\bf 1}+aA{\bf x}-{\bf x}.
\label{PR_matrix}
\eeq
In this system, centrality is generated uniformly across the network at each node (first term on the right). It is also destroyed at each node at a rate given by the amount of centrality at the node (last term on the right). An individual's centrality is increased by a flux from its neighboring nodes in proportion to their own centrality (middle term on the right). We shall refer to this system as Katz dynamics.

The steady-state solution coincides with (\ref{cent_eq}) when $0< a<1/\mu_1$ where $\mu_1$ denotes the largest eigenvalue of $A$ (which may not be unique). We know that for non-negative $A$, $\mu_1$ is non-negative and equal to the spectral radius $\rho(A)$\cite{Grantmacher}.

To determine the stability of the steady-state solution, we form the Jacobian of (\ref{PR_matrix}). This is $J=aA-I$, which has characteristic equation $det(J-\lambda I)=0$ or
\beq
det\left (A-\frac{\lambda+1}{a}I\right )=0,
\nonumber
\eeq
provided $a\neq 0$. The eigenvalue $\lambda_1$ of $J$ with the largest real part is therefore $\lambda_1=a\mu_1-1$ and it follows that the condition for stability of the steady state of (\ref{PR_matrix}) is $a<1/\mu_1$. This is also the condition for $I-aA$ to be an M-matrix\cite{Johnson} which implies that all elements of its inverse are non-negative; clearly this is a desirable property in forming a centrality measure. The stability of the steady state of~(\ref{PR_matrix}) is also essential for our control analysis below. 

\subsection*{Control analysis}
When the condition $a<1/\mu_1$ is met, (\ref{PR_matrix}) is a structurally stable continuous-time dynamical system describing dynamics on a network with matrix $A$ comprising $n$ variables $x_1,x_2,...,x_n$. We define the importance or centrality of a node by the net impact on the system of its absence\cite{Hadjichrysanthou}. So, the total impact of a node is the sum of the differences in the steady states of (\ref{PR_matrix}) before and after the node removal. This accounts both for the net effect of the flux passed to the node itself, but also the contribution it makes to others in the system.    

Define ${\bf q}^i=[q_1^i,q_2^i,...,q_{n-1}^i]^T$ to be the steady-state solution (\ref{cent_eq}) with node $i$ removed from $A$. Then define ${\bf r}^i=[q_1^i, q_2^i,...,$ $q_{i-1}^i,0, q_i^i, q_{i+1}^i,..., q_{n-1}^i]^T$; that is, each node in the steady-state solution of the perturbed network maintains the same position in the vector ${\bf r}^i$ as in ${\bf r}$ and $r_i^i=0$. Then we can define the total impact of node $i$:
\beq
d_i={\bf 1}^T({\bf r}-{\bf r}^i).
\label{deletion}
\eeq
This needs to be repeated for each node in the network to form the full vector of centralities ${\bf d}=[d_1,d_2,...,d_n]^T$. It is important to observe that node deletion cannot lead to an increase in the principal eigenvalue\cite{Grantmacher} and so the stability condition is never broken.

The computational complexity of assessing the impact of perturbations on this linear system is not generally reducible\cite{Skeel} and typically requires that we solve as many linear systems of equations as there are nodes in the network. While we shall compare our numerical results with this direct deletion method, we shall now derive an approximation to this based on a linear sensitivity analysis similar to metabolic control analysis\cite{Kacser,Heinrich,Reder} used in the analysis of biochemical networks. This reduces computational cost and yields some new perspectives.

We consider small perturbations to the steady state of (\ref{PR_matrix}) by targeting individual nodes. The resulting linear response of the steady states of all other nodes is then used to approximate the process of node deletion. To enable targeting of individual nodes, we rewrite (\ref{PR_matrix}) to introduce some node-specific parameters ${\boldsymbol{\gamma}}=[\gamma_1,\gamma_2,...,\gamma_n]^T$:
\beq
\frac{d{\bf x}}{dt}={\bf 1}+aA{\bf x}-{\boldsymbol{\gamma}} \circ {\bf x},
\label{PR_matrix2}
\eeq
where $\circ$ denotes the Hadamard (or component-wise) product. Now define
\beq
{\bf f}({\bf x},{\boldsymbol{\gamma}})={\bf 1}+aA{\bf x}-{\boldsymbol{\gamma}} \circ {\bf x}
\nonumber
\eeq
 to be a vector-valued function of variables ${\bf x}$ and ${\boldsymbol{\gamma}}$. Its zero values (${\bf f}({\bf x^*},{\boldsymbol{\gamma}})={\bf 0}$, where {\bf 0} is the column vector $[0,0,...,0]^T$ of length $n$) define the steady state ${\bf x^*}=[x^*_1,x^*_2,...,x^*_n]^T$ of the dynamical system (\ref{PR_matrix2}).

We shall assume that ${\boldsymbol{\gamma}}={\bf 1}$ because this is equivalent to the original system~(\ref{PR_matrix}), and in this case ${\bf x^*}={\bf r}$. However we use this new vector of parameters to investigate small node-specific perturbations from this steady state. Specifically, we define the impact of one node $i$ on another node $j$ by:
\beq
C_{ji}=\frac{dx_j^*}{d\gamma_i}\frac{\partial\gamma_i}{\partial x_i^*}x_i^*,
\label{eq5}
\eeq
where $\gamma_i$ provides a direct instantaneous perturbation to just the value $x_i^*$, followed by more complicated effects through the network as the system moves to a new state. Here, the linear response of the steady state to a change in $\gamma_i$ is given by $dx_j^*/d\gamma_i$. The remaining factor determines the size of perturbation equivalent to the removal of node $i$ and together they form a linear approximation to the removal of node $i$. 

To determine (\ref{eq5}), we first solve (\ref{PR_matrix2}) at the steady state for $x_i^*$:
\beq
x_i^*=\frac{1}{\gamma_i}\left [1+a\sum_jA_{ij}x_j^*\right ],
\nonumber
\eeq
from which we obtain:
\beq
\frac{\partial x_i^*}{\partial\gamma_i}=-\frac{x_i^*}{\gamma_i}.
\nonumber
\eeq
So, the immediate impact of a small increase in $\gamma_i$ is that $x_i^*$ reduces in proportion to its size.

For ${\boldsymbol{\gamma}}={\bf 1}$ we have
\beq
C_{ji}=-\frac{dx_j^*}{d\gamma_i}.
\label{eq5.1}
\eeq
After a small perturbation, the system returns to a new steady state near to the original one. Since ${\bf f}={\bf 0}$ at both steady-state solutions, an infinitesimal perturbation between two steady states is described by
\beq
\frac{d{\bf f}}{d{\boldsymbol{\gamma}}}={\bf 0_{n\times n}},
\nonumber
\eeq
where ${\bf 0_{n\times n}}$ is the $n$ by $n$ zero matrix. The total derivative of ${\bf f}$ with respect to ${\boldsymbol{\gamma}}$ is
\beq
\frac{d{\bf f}}{d{\boldsymbol{\gamma}}}=\frac{\partial {\bf f}}{\partial{\boldsymbol{\gamma}}}+\frac{\partial {\bf f}}{\partial {\bf x}}\frac{d{\bf x}}{d{\boldsymbol{\gamma}}}.
\nonumber
\eeq
Putting $d{\bf f}/d{\boldsymbol{\gamma}}={\bf 0_{n\times n}}$ and determining the relevant partial derivatives enables us to obtain $C$ from (\ref{eq5.1}):
\begin{eqnarray}\nonumber
C=-\frac{d{\bf x^*}}{d{\boldsymbol{\gamma}}}&=&\left [\left.\frac{\partial {\bf f}}{\partial {\bf x}}\right|_{({\bf r},{\bf 1})}\right ]^{-1}\left [\left.\frac{\partial {\bf f}}{\partial{\boldsymbol{\gamma}}}\right|_{({\bf r},{\bf 1})}\right ] \\ \nonumber
&=&\left [aA-I\right ]^{-1}\left [-R\right ] \\
&=& MR,
\nonumber
\end{eqnarray}
where $R$ is a diagonal matrix with diagonal elements $R_{ii}=r_i$. 

The element $C_{ji}$ describes the impact of node $i$ on node $j$. We can construct the total impact on the system of perturbing node $i$ by computing $\sum_jC_{ji}$, and for all nodes, this can be determined by taking the column-sum of $C$:
\beq
\boldsymbol{\sigma}=({\bf 1}^TMR)^T=RM^T{\bf 1}.
%\label{control_centrality}
\nonumber
\eeq
Using (\ref{cent_eq}) and the Hadamard product we obtain
\beq
\boldsymbol{\sigma}=(M{\bf 1})\circ (M^T{\bf 1}),
\label{LCA}
\eeq
which is the component-wise product of the row-sum and the column-sum of $M$ which we shall refer to here as the linear approximation.

The row-sum of $M$ corresponds to Katz centrality. The utility of the column-sum of $M$ in describing the influence of nodes, or what we shall refer to as the `sender' property, has been emphasised elsewhere \cite{Taylor}. Hence, using our knowledge of ${\bf r}$ as a measure of the hub-like or `receiver' property gives the centrality $\boldsymbol{\sigma}$ as a product of measures of the extent to which a node is a sender or a receiver:
\beq
\boldsymbol{\sigma}={\bf r}\circ {\bf s},
\label{K}
\eeq
where
\beq
{\bf s}=M^T{\bf 1}
\label{sender}
\eeq
is a vector ${\bf s}=[s_1,s_2,...,s_n]^T$ denoting the sender property of each node. It is straightforward to see that ${\bf s}$ is also the steady-state solution to Katz dynamics on $A^T$.
 
\subsection*{Results}
\subsection*{General observations}
By making a linear approximation to the process of removing nodes from the steady state of Katz dynamics we obtain a product of two quantities, one quantifying the `sender' property of a node and one quantifying the `receiver' property. For our overall measure of centrality $\boldsymbol{\sigma}$ to be large, we typically require both the sender (${\bf s}$) and receiver (${\bf r}$) properties to be large. 

We note that ${\bf r}$ and ${\bf s}$ describe very similar network properties to Kleinberg's hubs (${\bf h}$) and authorities (${\bf a}$) respectively and also to the principal right (${\bf u}$) and principal left (${\bf v}$) eigenvectors of $A$ respectively. Motivated by (\ref{K}), it is also of interest to numerically investigate the analogous products ${\bf h}\circ {\bf a}$ and ${\bf u}\circ{\bf v}$ which we shall do in the following section.

The form of (\ref{LCA}) means that $\boldsymbol{\sigma}$ is invariant with respect to taking the transpose of $A$. Furthermore the sender and receiver properties are interchanged under this operation. As a special case of this, on undirected networks it follows that ${\bf s}={\bf r}$ and that consequently for each node $i$, $\sigma_i=r_i^2=s_i^2$. Due to this monotonic relationship, the node rankings for undirected networks are the same for $\boldsymbol{\sigma}$ as for Katz centrality;  node rankings only differ on networks with directed links. It also follows from the definitions of the sender and receiver properties that, over the whole network, what is sent is also received; ${\bf 1^Tr}={\bf 1^Ts}$. So in this sense, the centrality flux is conserved. Analogous results to these also apply for ${\bf h}\circ {\bf a}$ and ${\bf u}\circ{\bf v}$.

For the original node deletion process, the form of (\ref{deletion}) means that ${\bf d}$ is also invariant with respect to taking the transpose of $A$, although in this case we did not propose an explicit separation into hub-like and authority-like properties. 

\subsection*{Numerical evaluation on example networks}
\begin{figure}
\centerline{\includegraphics[width=.7\linewidth]{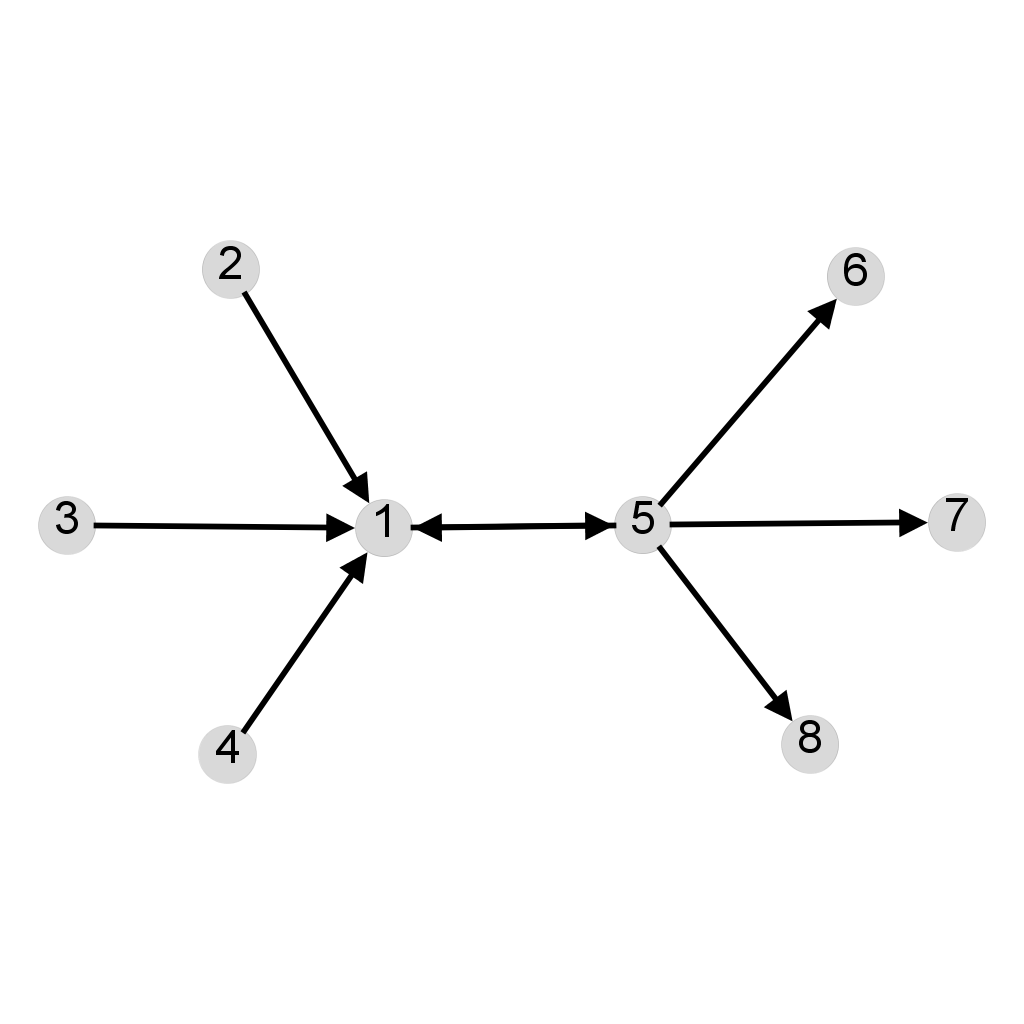}}
\caption{A simple example network emphasising the effect of peripheral source and sink nodes and the symmetry of node deletion under matrix transposition.}
\label{Synthetic_graph}
\end{figure}

Table~\ref{Synth_tab} shows the different centrality measures that we have considered applied to the example network in Figure~\ref{Synthetic_graph}. For Katz dynamics, the impact of the deletion of the peripheral source and sink nodes is captured exactly by the linear approximation. Additionally, nodes 1 and 5 have the same centrality under the node deletion process and this is also captured qualitatively by the linear approximation, although the numerical values are quite different. The equal centrality of nodes 1 and 5 can be understood in terms of the invariance with respect to the matrix transpose discussed above. Observe that due to the structure of this network, nodes 1 and 5 interchange roles under the transpose of $A$; this, together with the invariance under this operation, implies that they must have the same value.

For the hub and authority measures of Kleinberg, we see from Table~\ref{Synth_tab} that each node acts as a pure hub or pure authority on this network. Consequently the product of these quantities is zero for each node and does not yield information about their relative importance. However, we could argue that nodes 1 and 5 act as a through-routes for the other nodes and in this sense, could usefully be allocated a centrality that acknowledges this. For the centrality formed by the product of the left and right principal eigenvectors, nodes 1 and 5 are given non-zero centrality. Again, the peripheral nodes are not given any centrality and are classed as pure source or sink nodes. Arguably this is reasonable, however, on other directed networks, nodes which have a role linking one node to another can also have zero eigenvector centrality\cite{Newman}.

\begin{table}
\caption{Sender (${\bf s}$), receiver (${\bf r}$), linear approximation ($\boldsymbol{\sigma}$) and node deletion (${\bf d}$) values are shown for the network in Figure~\ref{Synthetic_graph}. Results for hubs (${\bf h}$), authorities (${\bf a}$), left eigenvector (${\bf v}$) and right eigenvector (${\bf u}$) are also shown, along with the products (${\bf h}\circ {\bf a}$) and (${\bf u}\circ{\bf v}$). For Katz dynamics, we used $a=0.85/\mu_1$ where $\mu_1$ is the largest eigenvalue of matrix $A$. Eigenvectors are scaled by the Euclidean norm, but no scaling is applied to ${\bf s}$ or ${\bf r}$ to enable a direct comparison with ${\bf d}$.}
\begin{center}
\begin{tabular}{|c|cccc|ccc|ccc|}\hline
Node & \multicolumn{4}{c|}{Katz dynamics} & \multicolumn{3}{c|}{Hubs/Authorities} &\multicolumn{3}{c|}{Eigenvectors} \\
&  ${\bf r}$ (Katz) & ${\bf s}$ & $\boldsymbol{\sigma}$ (LCA)& ${\bf d}$ (deletion) &${\bf h}$  & ${\bf a}$ & ${\bf h}\circ{\bf a}$ & ${\bf u}$ (Right) & ${\bf v}$ (Left)& ${\bf u}\circ{\bf v}$\\ \hline
       1 &     15.9 &    14.5 &     230 &     63.7 &     0.799 &    0 &     0 &     0.447 &    0.447 &     0.2 \\
			2-4 & 1 & 13.3 & 13.3 &13.3 & 0 & 0.347 & 0 & 0 & 0.447 & 0 \\
			5 & 14.5 & 15.9 & 230 & 63.7 & 0 & 0.799 & 0 & 0.447 & 0.447 & 0.2 \\
			6-8 & 13.3 & 1 & 13.3 & 13.3 & 0.347 & 0 & 0 & 0.447 & 0 & 0 \\ \hline
			\end{tabular}
\end{center}
\label{Synth_tab}
\end{table}

\begin{figure}
\centerline{\includegraphics[width=.9\linewidth]{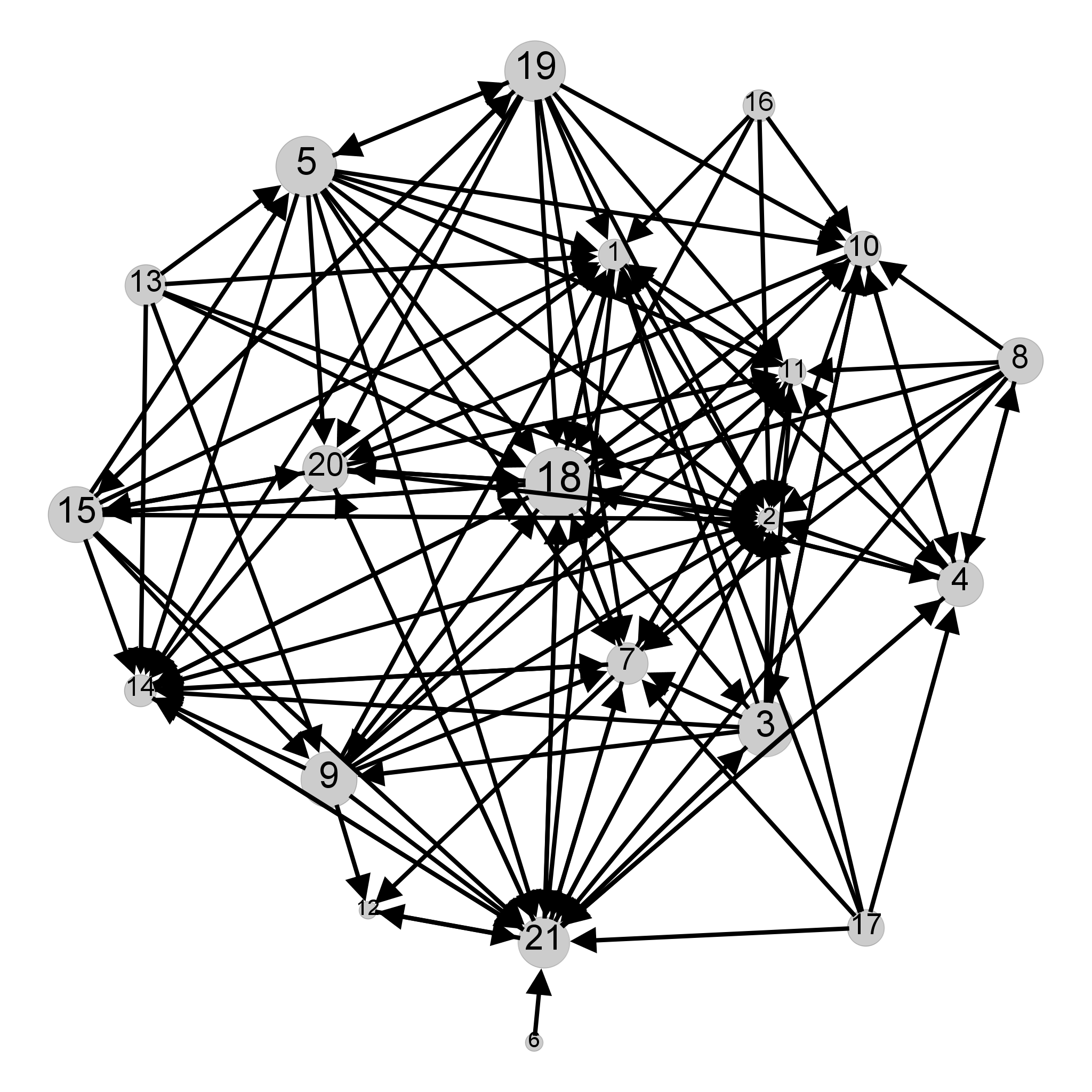}}
\caption{The social interaction network of Krackhardt\cite{Krackhardt} depicting advice structures within an organisation. The in-degree is easily seen from the number of arrow heads. The size of nodes is indicative of their out-degree (see Table~\ref{Krackhardt_results_table} for actual in-degrees and out-degrees). This data is obtained by interviewing 21 members of an organisation and asking each one about how they perceive management or advice structures between all individuals. Here we use the network termed by Krackhardt as the Locally Aggregated Structure (LAS) formed from all of the links where both individuals at either end of a link agree that the link exists. A link from an individual $i$ to an individual $j$ indicates that $i$ goes to $j$ for help and advice.}
\label{Krackhardt}
\end{figure}

Figure~\ref{Krackhardt} shows a directed network
  describing the advice structures between individuals in an organisation based on data compiled by Krackhardt\cite{Krackhardt}. The rankings of each node in this network are given in Table~\ref{Krackhardt_results_table}. For comparison, the in-degree and out-degree of each node is also given. The ranking with respect to hubs/authorities, right/left eigenvectors and receiver/sender have a high degree of consistency, as we would expect because node degree remains the main driver of each of these, but there are also significant differences.

\begin{table}
\caption{Ranking of nodes on the Krackhardt LAS Network (Figure~\ref{Krackhardt}). Nodes are ranked from 1 (most important) to 21 (least important). Nodes with the same ranking occur when the method yields a centrality of zero (or of 1 in the case of Katz centrality). The in-degree and out-degree of nodes are provided for comparison. For Katz dynamics, we used $a=0.85/\mu_1$ where $\mu_1$ is the largest eigenvalue of the adjacency matrix of the network.  }
\begin{center}
\begin{tabular}{|c|cc|cccc|ccc|ccc|}\hline
Node & \multicolumn{2}{c|}{Degree} & \multicolumn{4}{c|}{Katz dynamics (Rank)} & \multicolumn{3}{c|}{Hubs/Authorities (Rank)} &\multicolumn{3}{c|}{Eigenvectors (Rank)} \\
& In & Out & ${\bf r}$ (Katz) & ${\bf s}$ & $\boldsymbol{\sigma}$ (LA)& ${\bf d}$ (deletion) &${\bf h}$  & ${\bf a}$ & ${\bf h}\circ{\bf a}$ & ${\bf u}$ (Right) & ${\bf v}$ (Left)& ${\bf u}\circ{\bf v}$\\ \hline
  1 &    12 &     4 &     6 &     14 &     9 &     9 &     4 &    16 &     5 &     7 &    14 &     8\\
     2 &    18 &     2 &     1 &     18 &     8 &     6 &     1 &    19 &    11 &     1 &    18 &    10\\
     3 &     3 &     9 &    11 &      5 &     6 &     8 &    15 &     3 &    14 &    10 &     5 &     5\\
     4 &     6 &     7 &     8 &     10 &     4 &     4 &    10 &     9 &    10 &     8 &    10 &     4\\
     5 &     3 &    10 &    16 &      4 &    14 &    14 &    13 &     1 &    12 &    16 &     4 &    16\\
     6 &     0 &     1 &    18 &     20 &    21 &    21 &    18 &    20 &    18 &    18 &    20 &    18\\
     7 &    11 &     6 &     4 &     13 &     3 &     3 &     6 &    11 &     3 &     4 &    13 &     3\\
     8 &     1 &     7 &    15 &      9 &    16 &    16 &    17 &     8 &    16 &    15 &     9 &    14\\
     9 &     4 &     9 &    13 &      7 &    12 &    12 &    11 &     4 &     7 &    13 &     7 &    12\\
    10 &     8 &     5 &    10 &     12 &    11 &    11 &     8 &    17 &     8 &    11 &    12 &    11\\
    11 &     9 &     3 &     7 &     19 &    13 &    13 &     7 &    18 &     9 &     6 &    19 &    13\\
    12 &     3 &     1 &    12 &     20 &    17 &    17 &    14 &    20 &    17 &    12 &    20 &    17\\
    13 &     0 &     6 &    18 &     11 &    18 &    18 &    18 &    12 &    18 &    18 &    11 &    18\\
    14 &    10 &     4 &     5 &     16 &     7 &     7 &     5 &    14 &     4 &     5 &    15 &     7\\
    15 &     3 &     9 &    14 &      2 &    10 &    10 &    12 &     6 &    13 &    14 &     2 &     9\\
    16 &     0 &     4 &    18 &     17 &    20 &    20 &    18 &    15 &    18 &    18 &    17 &    18\\
    17 &     0 &     5 &    18 &     15 &    19 &    19 &    18 &    13 &    18 &    18 &    16 &    18\\
    18 &    15 &    12 &     3 &      1 &     1 &     1 &     2 &     2 &     1 &     3 &     1 &     1\\
    19 &     2 &    10 &    17 &      3 &    15 &    15 &    16 &     5 &    15 &    17 &     3 &    15\\
    20 &     6 &     7 &     9 &      8 &     5 &     5 &     9 &     7 &     6 &     9 &     8 &     6\\
    21 &    15 &     8 &     2 &      6 &     2 &     2 &     3 &    10 &     2 &     2 &     6 &     2\\ \hline
\end{tabular}
\end{center}
\label{Krackhardt_results_table}
\end{table}

Neither Katz centrality, nor the products ${\bf a}\circ {\bf h}$ and ${\bf u}\circ {\bf v}$ distinguish between the nodes 6,13,16 and 17, ranking them all as least important. However, for many purposes we would consider, for example, node 6, which only connects to node 21, to be less influential than node 17 which connects to five other individuals including node 21. Katz centrality and right-eigenvector centrality favour nodes which have large in-degree and this is particularly apparent in the top-ranking of node 2, in spite of it having a relatively low out-degree of only 3, meaning that it is close to resembling a sink. Node 18 is clearly very central in the sense of having the largest out-degree and a large in-degree and is identified as the most central node by the direct deletion method, the linear approximation and the combinations ${\bf h}\circ {\bf a}$ and ${\bf u}\circ{\bf v}$.

\begin{figure}
\centerline{\includegraphics[width=.6\linewidth]{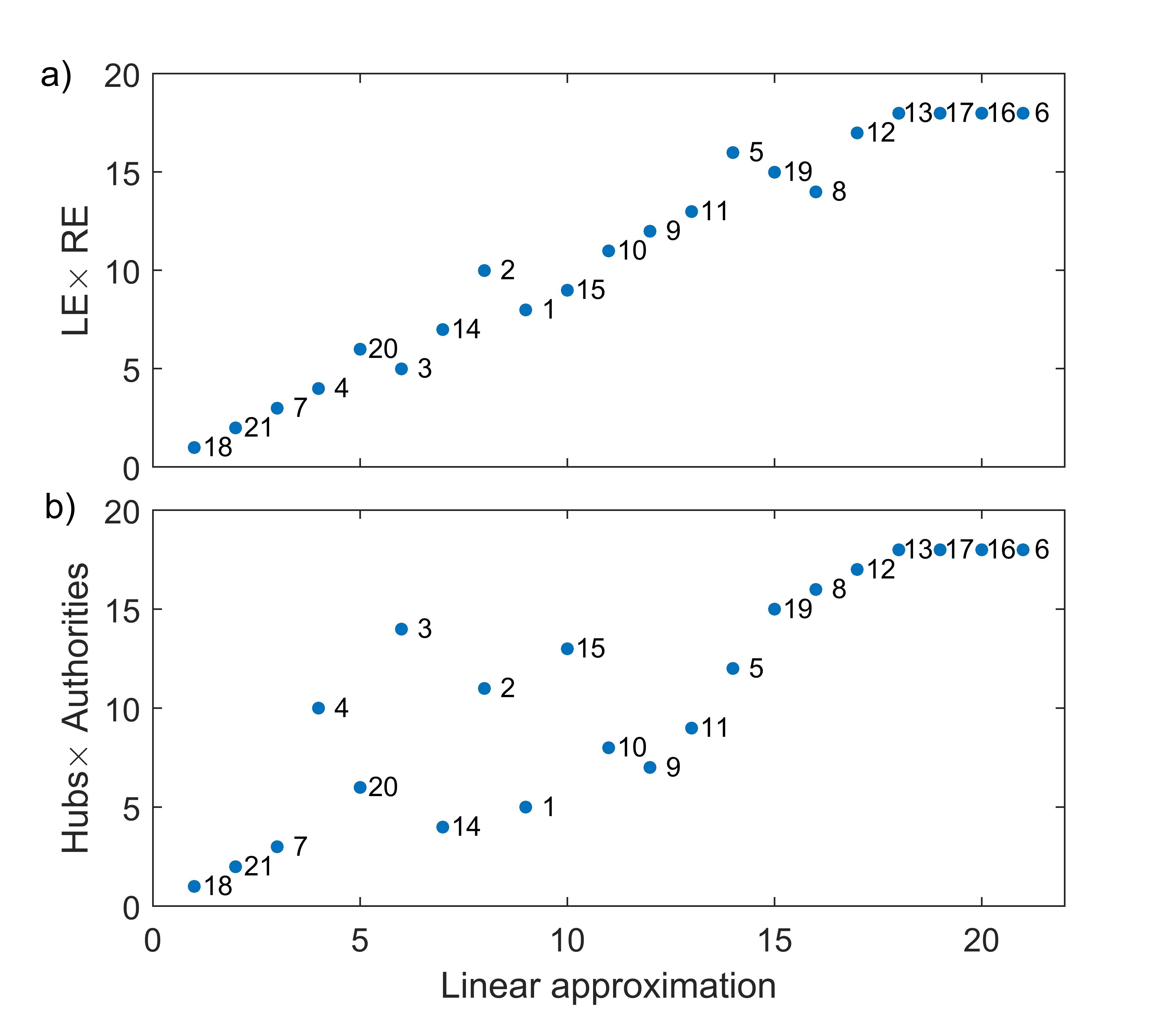}}
\caption{Node rankings on the Krackhardt network (Figure~\ref{Krackhardt}) given by a) the product of left and right principal eigenvectors and b) the product of hubs and authorities plotted against the node rankings from the linear approximation (${\boldsymbol{\sigma}}$). For Katz dynamics we used $a=0.85/\mu_1$ where $\mu_1$ is the largest eigenvalue of the adjacency matrix. Node identifiers correspond to those in Figure~\ref{Krackhardt}.}
\label{scatterplot2}
\end{figure}

Figure~\ref{scatterplot2} gives a visualisation of some of the information in Table~\ref{Krackhardt_results_table}. Here the rankings given by ${\bf h}\circ {\bf a}$ and ${\bf u}\circ{\bf v}$ are plotted against those given by the linear approximation ($\boldsymbol{\sigma}$) since these are the two most directly comparable quantities.
We have already noted the significant feature that nodes 6,13,16 and 17 are not differentiated by either ${\bf h}\circ {\bf a}$ or ${\bf u}\circ{\bf v}$, but they are ranked by $\boldsymbol{\sigma}$. From Figure~\ref{scatterplot2}a, it is clear that the rankings of the linear approximation are similar to ${\bf u}\circ{\bf v}$ on this network, reflecting the close relationship between eigenvector centrality and Katz centrality\cite{Newman}. More significant differences are apparent in the rankings by hubs and authorities in Figure~\ref{scatterplot2}b.

\begin{figure}
\centerline{\includegraphics[width=.6\linewidth]{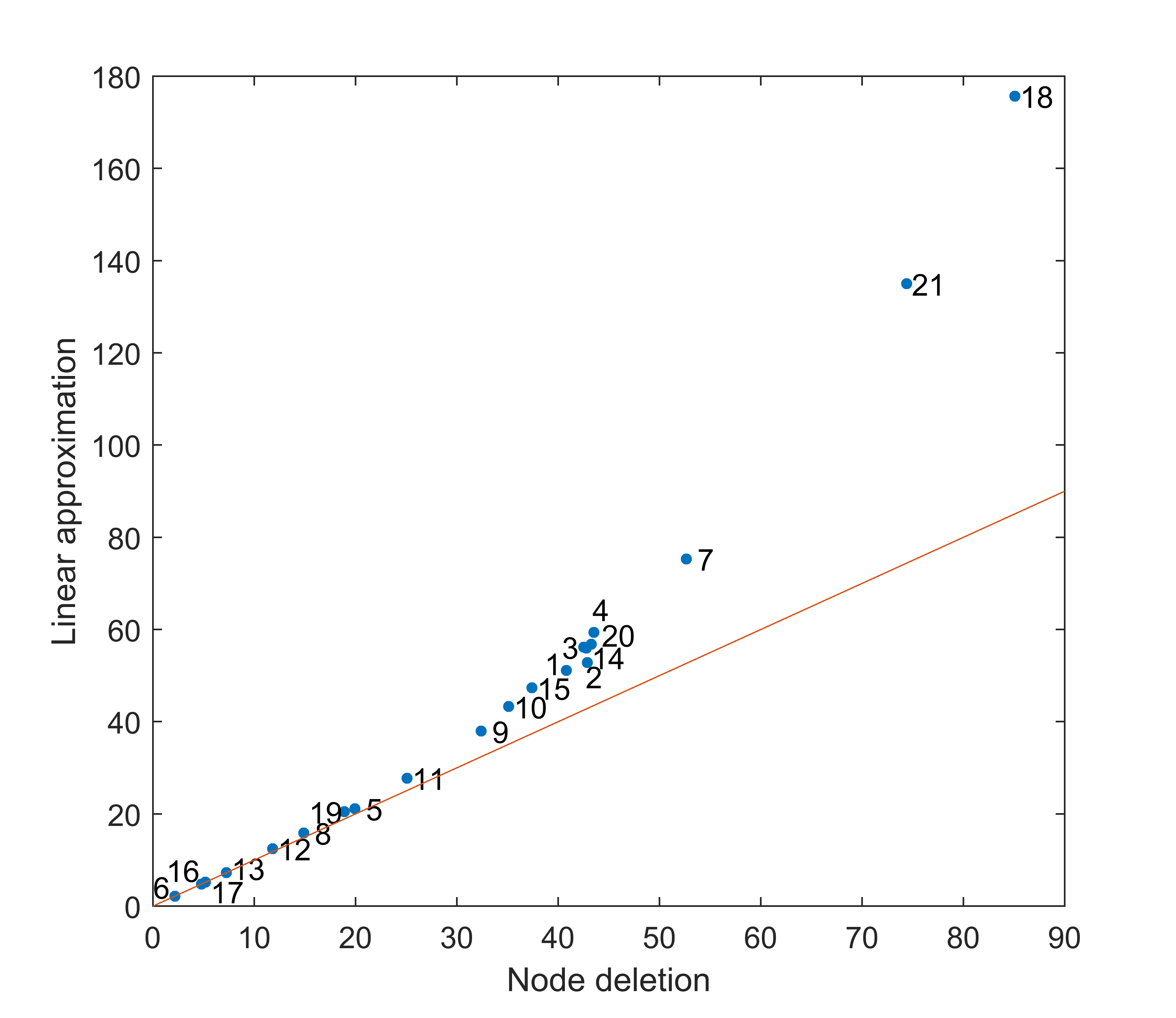}}
\caption{Plot of the linear approximation ($\boldsymbol{\sigma}$) against direct deletion (${\bf d}$) for the Krackhardt network (Figure~\ref{Krackhardt}). No scaling is applied to these values to enable a direct comparison. The line of equality is also plotted to assist in seeing the deviation. Here we used $a=0.85/\mu_1$ where $\mu_1$ is the largest eigenvalue of the adjacency matrix of the network. Node identifiers correspond to those in Figure~\ref{Krackhardt}.}
\label{scatterplot1}
\end{figure}

Recalling that $\boldsymbol{\sigma}$ is obtained as a linear approximation to the node deletion process, it is of interest to investigate how accurate this approximation is. For the network in Figure~\ref{Synthetic_graph}, some comparisons between ${\bf d}$ and its linear approximation $\boldsymbol{\sigma}$ are made in Table~\ref{Synth_tab}. For the Krackhardt network (Figure~\ref{Krackhardt}), a comparison between the values of ${\bf d}$ and the linear approximation is given in Figure~\ref{scatterplot1}.
On this network, the linear approximation appears to accurately quantify the deletion process for the lower centrality nodes, but deviates for the higher-centrality nodes where the perturbation is larger. However, the almost monotonic relationship between the two means that the rankings of the nodes are almost identical, as can be seen in Table~\ref{Krackhardt_results_table}.

\section*{Discussion}
\label{S6}
We introduced a new interpretation of Katz centrality as the unique steady-state solution of an appropriate continuous-time dynamical system. We argued that by removing a node from the network and investigating the net impact on this steady state, both the direct value of the node as well as the impact that the node has on others is determined. This contrasts with some standard centrality measures such as Katz centrality and eigenvector centrality which only quantify the direct value of the node rather than its contribution to the whole system.

A linear approximation to this deletion process yielded a new centrality which is a product of two quantities, one of which quantifies the capacity of a node to receive flux from its neighbours (the receiver property) and the other which represents the capacity of a node to pass flux on to its neighbours (the sender property). The receiver property (\ref{cent_eq}) is equal to the original Katz centrality and is formed from the row-sum of matrix $M$. The sender quantity (\ref{sender}) corresponds to the column-sum of $M$ and has been highlighted before as useful for capturing the influence of nodes on others\cite{Taylor}. In this sense, the centrality formed by the product of both requires that a central node is one that is able to receive flux and then pass it on to others and thereby contribute to the ongoing dynamics. Nodes that act just as sources or as sinks are not central to propagating and maintaining the dynamics of the system and in this sense are peripheral, resulting in a low overall centrality score. However, their separate sender or receiver properties could be large.

For the purposes of comparison, the form of the linear approximation to the node deletion process led us to define analogous measures given by the product of Kleinberg's hubs and authorities and from the product of the left and right principal eigenvectors of the adjacency matrix. As expected, there is some correlation between these measures and with the linear approximation. However, a problem with both of these eigenvector methods is that they frequently yield zero centrality for some nodes on directed networks and so give no information on their relative importance. To determine eigenvector centrality unambiguously and with no zero values, we require that the adjacency matrix $A$ is irreducible (strongly connected). For Kleinberg's hubs and authorities, zero values can occur unless the matrices $AA^T$ and $A^TA$ are irreducible. On directed networks this requirement is frequently broken, such as for both of our example networks.

In contrast, when defining the node deletion process ${\bf d}$ and its linear approximation $\boldsymbol{\sigma}$, we only require that the matrix $I-aA$ is non-singular, which we ensure with a suitable choice for parameter $a$. So, both the sender and receiver properties are non-zero and their product gives non-zero centralities for all nodes on any network, which is arguably advantageous. Indeed, Katz centrality itself has been viewed as a modification of right-eigenvector centrality to give each node an intrinsic self-centrality\cite{Newman} to resolve some of the problems on directed networks. Similarly, the sender property that we defined can be regarded as an equivalent modification of left-eigenvector centrality.  

A natural question that arises is whether our original node deletion definition of centrality (${\bf d}$) can usefully be decomposed into sender and receiver properties in a similar way to its linear approximation. Other directions may include higher-order approximations to node deletion, a control analysis approach to modifications of Katz dynamics such as PageRank, or to a Katz-like variant of the hub and authority eigenvector equations\cite{Newman}.

\section*{Acknowledgements}
This work was funded by the Leverhulme Trust Research Project Grant RPG-2014-341.

\end{document}